\documentclass[twocolumn, numberedappendix, twocolappendix, apj, iop, unicode, 
colorlinks=true,
linkcolor=linkcolor,
citecolor=linkcolor,
filecolor=linkcolor,
urlcolor=linkcolor,
hyperfootnotes=true]{openjournal}
\usepackage{amsmath,amsfonts,amssymb}
\usepackage[caption=false]{subfig}
\usepackage[dvipsnames,usenames]{xcolor}
\usepackage{ulem}
\usepackage{fancyhdr}
\usepackage{comment}
\usepackage{ragged2e}

\usepackage{epsfig}
\usepackage{graphicx}
\usepackage{bm}
\usepackage{float} 
\usepackage{color}
\usepackage{orcidlink}
\usepackage{dcolumn}
\usepackage{booktabs}

\renewcommand{\doi}[1]{\href{https://doi.org/#1}{DOI: \nolinkurl{#1}}}

\newcommand{\dd}{\mathrm{d}}
\newcommand{\ket}[1]{\left\vert{#1}\right\rangle}
\newcommand{\bra}[1]{\left\langle#1\right\vert}
\newcommand{\alphaC}{\alpha_{\text{C}}}
\newcommand{\alphaG}{\alpha_{\text{G}}}
\newcommand{\betaC}{\beta_{\text{C}}}
\newcommand{\betaG}{\beta_{\text{G}}}

\definecolor{linkcolor}{rgb}{0.0,0.3,0.5}

\begin{document}

\title{Primordial magnetogenesis in a bouncing model with dark energy}

\author{Marcus V. Bomfim}
\email{marcusbomfimdejesus@gmail.com}
\affiliation{Centro Brasileiro de Pesquisas Físicas, Rua Dr.~Xavier Sigaud 150,
    Urca, CEP 22290-180, Rio de Janeiro, RJ, Brazil}

\author{Emmanuel Frion~\orcidlink{0000-0003-1280-0315}}
\email{efrion@uwo.ca}
\affiliation{Department of Physics and Astronomy, Western University,
    N6A 3K7, London, Ontario, Canada}

\author{Nelson Pinto-Neto~\orcidlink{0000-0001-6713-5290}}
\email{nelsonpn@cbpf.br}
\affiliation{Centro Brasileiro de Pesquisas Físicas, Rua Dr.~Xavier Sigaud 150,
    Urca,  CEP 22290-180, Rio de Janeiro, RJ, Brazil}
\affiliation{PPGCosmo, CCE, Universidade Federal do Espírito Santo, Vitória,
    29075-910, Espírito Santo, Brazil}

\author{Sandro D. P. Vitenti~\orcidlink{0000-0002-4587-7178}}
\email{vitenti@uel.br}
\affiliation{Departamento de Física, Universidade Estadual de Londrina, Rod. Celso
    Garcia Cid, Km 380, 86057-970, Londrina, Paraná, Brazil}


\begin{abstract}
    We investigate primordial magnetogenesis within a quantum bouncing model driven by a
    scalar field, focusing on various non-minimal couplings between the electromagnetic
    field and the scalar field. We test three cases: no coupling, a Cauchy coupling with gradual decay, and a Gaussian coupling with rapid fall-off. By exploring these
    scenarios, we assess a wide range of coupling strengths across different scales. The
    scalar field, with an exponential potential, behaves as pressureless matter in the
    asymptotic past of the contracting phase, as stiff matter around the bounce, and as
    dark energy during the expanding phase. Our findings reveal that, among the tested
    cases, only the Gaussian coupling can explain the generation of primordial magnetic
    fields on cosmological scales.
\end{abstract}

\maketitle

\section{Introduction}

Bouncing models provide an excellent framework for investigating gravitational particle
creation, as they necessarily include a time period characterized by strong
gravitational fields. Such models can produce scalar and fermionic particles
\cite{bib:Pinto-Neto2017, bib:Brandenberger2014, bib:Pinto-Neto2018}, baryons \cite{bib:Pinto-Neto2020}, and primordial
magnetic fields (PMF) \cite{Frion:2020bxc, Battefeld:2004cd, Salim:2006nw, Membiela2013,
    Sriramkumar:2015yza, Chowdhury2016, Qian2016, Koley:2016jdw, Chen:2017cjx,
    Leite:2018bbo, Chowdhury:2018blx, Barrie:2020kpt, Motaharfar:2024afn}, which may serve
as seeds for the magnetic fields observed across various scales \cite{Durrer2013,
    beck2012magnetic, Beck:2013bxa, Minoda:2018gxj, Bray:2018ipq, Ade:2015cva,
    Chluba:2019kpb, Zucca:2016iur, Pogosian:2018vfr, Saga:2017wwr, Kawasaki:2012va,
    Barai:2018msb}.

The creation of primordial magnetic fields has also been explored within the context of
non-minimal couplings between electromagnetic fields and scalar fields in a variety of
models \cite{Emami2009, Adshead:2016iae, Turner1988, Bamba2006, Campanelli2008,
    Kunze2009, Kunze2012, Savchenko2018}. This paper aims, as a sequence of
Ref.~\cite{Frion:2020bxc}, to examine the possibility of producing PMF arising from these
electromagnetic-scalar field non-minimal couplings within the framework of a bouncing
model. To this end, we adopt the background bouncing model introduced in
Ref.~\cite{Bacalhau:2017hja}. This model features a scalar field with an exponential
potential, which is both simple and rich in its implications, as the scalar field
behaves as pressureless matter in the asymptotic past of the contracting phase, as stiff
matter around the bounce, and as dark energy during a portion of the expanding phase.
Furthermore, this model produces scalar and tensor perturbation spectra and amplitudes
compatible with observations of the Cosmic Microwave Background (CMB). Notably, quantum
effects at the bounce lead to the magnification of scalar perturbations relative to
tensor perturbations, a feature that is challenging to achieve with canonical scalar fields,
at least within the classical realm.

The non-minimal couplings investigated between the electromagnetic and scalar fields
include Gaussian and Cauchy functions of the scalar field. These couplings vary
significantly in their effectiveness: Gaussian couplings exhibit a rapid fall-off, while
Cauchy couplings decay more slowly, extending their influence over a broader range. By
studying both fast-decaying (Gaussian) and slow-decaying (Cauchy) types, our analysis
not only further explores the Gaussian coupling but also provides a comprehensive
examination of the effects of different coupling decay rates. While Gaussian couplings
have been used in inflationary magnetogenesis scenarios \cite{Tripathy:2021sfb}, their
application to bouncing magnetogenesis has not been explored previously.

We divide the paper as follows: In Sec.~\ref{backg}, we describe the background model,
including the scalar and gravitational fields. In Sec.~\ref{emsector}, we outline the
electromagnetic sector and describe its quantization and vacuum initial conditions. In
Sec.~\ref{numeric}, we present the two non-minimal couplings studied and perform
numerical calculations to obtain the amplitudes of the electric and magnetic field,
power spectra, and spectral indices. We also address back-reaction and make comparisons
with observations. We conclude and discuss our findings in Sec.~\ref{discussion}.

\section{The background}
\label{backg}

In this section, we summarize the minisuperspace canonical quantization of a
cosmological model featuring flat, homogeneous, and isotropic spacelike hypersurfaces, coupled with a canonical scalar field that has an exponential potential $V(\phi)=V_0 \exp(\lambda\kappa\phi)$, 
where $\kappa\equiv 1/M_p\equiv\sqrt{8\pi G_N}$, $G_N$ is the Newton constant, and $\lambda,V_0$ are free parameters. It is well known that, in this case, all scalar field solutions of the equations of motion in the asymptotic past of the contracting phase behave as a perfect fluid with $p=w\rho$ and $w=(\lambda^2-3)/3$, see \cite{Heard:2002}. We choose $\lambda$ such that $w\approx 0$, as it is well known that in this case, scalar cosmological perturbations originating from vacuum quantum fluctuations in a dust-dominated cosmological asymptotic past of such bouncing models present an almost scale-invariant spectrum of scalar perturbations, as observed, see \cite{Peter:2008qz}. 

As the contracting
phase evolves, the kinetic term $\dot{\phi}^2$ increases substantially over the potential term, implying the the scalar field behaves like a stiff matter fluid $p=\rho$ when the model is approaching the classical singularity.
In this phase, we can
neglect the potential term for the purpose of quantization. This approximation allows us to use the results of Ref. \cite {colistete2000gaussian}, which shows that quantum effects can
prevent the cosmological singularity. Hence, the model contains a long
contracting period which transitions smoothly through a regular bounce to the expanding
phase observed today. In the expanding phase, the scalar field evolves to exhibit a
brief period during which it acts as a dark energy component. For further details, see
\cite{Bacalhau:2017hja}.

Let us focus on the bounce phase, during which quantum effects become relevant and the
potential energy density of the scalar field $\phi$ is negligible compared to its
kinetic energy density. This condition implies that $p=\rho$, corresponding to a ``stiff
matter'' equation of state. The line element is characterized only by the scale factor
$a(t)$ and is given by
\begin{align}\label{eq: Line element with epsilon curvature}
    \dd s^{2}=N^{2} \dd\tau^{2}-a^{2}(\tau)\left(\dd x^{2}+ \dd y^{2}+ \dd z^{2}\right) \;.
\end{align}
Defining $\alpha =\ln{(a)}$, changing to the dimensionless scalar field $\phi\rightarrow \kappa\phi/\sqrt{6}$, and expressing the Planck length as $l_p \equiv
    \sqrt{G_{\mathrm{N}}}$, we can write the
Hamiltonian that governs the dynamics of the scalar field and scale factor as
\begin{equation}\label{eq: Hamiltonian RG Var ADM with Scalar Field a<->alpha.}
    H=N \mathcal{H}=\frac{N}{2 l_p \mathrm{e}^{3 \alpha}}\left(-\Pi_{\alpha}^{2}+\Pi_{\phi} ^{2}\right) \;.
\end{equation}
Here, $\Pi_{\alpha}$ and $\Pi_{\phi}$ are the canonically conjugated momenta to $\alpha$
and $\phi$, respectively:
\begin{subequations}
    \begin{align}
         & \Pi_{\alpha}=-\frac{l_{p}}{N}\mathrm{e}^{3 \alpha} \frac{\dd\alpha}{\dd\tau}, \\
         & \Pi_{\phi}=\frac{l_{p}}{N}\mathrm{e}^{3 \alpha} \frac{\dd\phi}{\dd\tau},
    \end{align}
\end{subequations}
The lapse function $N$ is a Lagrange multiplier, and the Hamiltonian is constrained to
vanish. In the Dirac quantization procedure for constrained systems \cite{Dirac:1950pj},
the constraints must annihilate the wave function, leading to
\begin{equation}
    \hat{\mathcal{H}} \Psi(\alpha, \phi)=0 \;.
\end{equation}
This condition results in the Wheeler-DeWitt equation of the model:
\begin{equation}\label{eq:Wheeler-DeWitt1}
    \hat{\mathcal{H}} \Psi(\alpha, \phi)=0 \Rightarrow \left[-\frac{\partial^{2}}{\partial \alpha^{2}}+\frac{\partial^{2}}{\partial \phi^2}\right] \Psi(\alpha, \phi)=0.
\end{equation}

Following Ref.~\cite{colistete2000gaussian}, we can obtain the quantum bouncing trajectories of the quantum cosmological model using the de Broglie-Bohm quantum theory, which must satisfy the so-called guidance equations (see reference \cite{Pinto-Neto:2013toa} for
motivations and details):
\begin{subequations}\label{eq:guide-eqs}
    \begin{align}
        \Pi_{\alpha} & =\frac{\partial S}{\partial \alpha}=-\frac{l_p}{N} \mathrm{e}^{3 \alpha} \frac{\dd\alpha}{\dd\tau}, \label{eq:guide-alpha} \\
        \Pi_{\phi}   & =\frac{\partial S}{\partial \phi}=\frac{l_p}{N} \mathrm{e}^{3 \alpha} \frac{\dd\phi}{\dd\tau}.\label{eq:guide-phi}
    \end{align}
\end{subequations}
The quantum potential, which incorporates quantum effects, is defined as
\begin{align}
    Q(\alpha, \phi)=\frac{e^{3 \alpha}}{2 R}\left[\frac{\partial^{2} R}{\partial \alpha^{2}}-\frac{\partial^{2} R}{\partial \phi^{2}}\right].
\end{align}

In the case of homogeneous and isotropic backgrounds, equations~\eqref{eq:guide-eqs} remain invariant under time reparametrization. Therefore, in this case,
even at the quantum level, different choices of $N(\tau)$ produce the same space-time
geometry for a given non-classical solution.

Equation~\eqref{eq:Wheeler-DeWitt1} represents a simple Klein-Gordon equation in
minisuperspace, and we can express the general solution as
\begin{equation}
    \begin{split}
        \Psi(\alpha, \phi) & = \int F(\bar{k}) \exp[i \bar{k}(\alpha-\phi)] \dd\bar{k}   \\
                           & + \int G(\bar{k}) \exp[i \bar{k}(\alpha +\phi)] \dd\bar{k}.
    \end{split}
\end{equation}
As established in Ref.~\cite{colistete2000gaussian}, we choose the Gaussian superposition
\begin{align}
    F(\bar{k})=G(\bar{k})=\exp \left[-\frac{(\bar{k}-d)^{2}}{\sigma^{2}}\right] ,
\end{align}
which yields the expression
\begin{align}
    \begin{aligned}
        \label{wave function}
        \Psi = & \sigma \sqrt{\pi}\left\{\exp\left[-\frac{(\alpha+\phi)^{2} \sigma^{2}}{4}\right] \exp [i d(\alpha+\phi)]\right. \\
               & \left.+\exp \left[-\frac{(\alpha-\phi)^{2} \sigma^{2}}{4}\right] \exp [-i d(\alpha-\phi) ]\right\}.
    \end{aligned}
\end{align}

We can extract the phase $S$ from Eq.~\eqref{wave function}. By inserting this phase
into the the guidance Eqs.~\eqref{eq:guide-eqs}, we obtain the Bohmian trajectories for
$\alpha (t)$ and $\phi(t)$, where $t$ is the cosmic time for which the lapse function is
$N=1$. The equations form a planar system represented as follows
\begin{subequations}\label{eq:bohm-planar}
    \begin{align}
         & l_p\dot{\alpha}=\frac{\phi \sigma^{2} \sin (2 d \alpha)+
            2 d \sinh \left(\sigma^{2} \alpha \phi\right)} {2 \mathrm{e}^{3 \alpha}\left[\cos (2 d \alpha)+
        \cosh \left(\sigma^{2} \alpha \phi\right)\right]}\; ,\label{eq:btrajalpha} \\
         & l_p \dot{\phi}=\frac{-\alpha \sigma^{2} \sin (2 d \alpha)+
            2 d \cos (2 d \alpha)+2 d \cosh \left(\sigma ^{2} \alpha \phi\right)}{2 \mathrm{e}^{3 \alpha}
            \left[\cos (2 d \alpha)+\cosh \left(\sigma^{2} \alpha \phi\right)\right]}.\label{eq:btrajphi}
    \end{align}
\end{subequations}
The dot denotes a derivative with respect to cosmic time $t$. This leads to the relation
\begin{equation}
    \begin{split}
         X &\equiv \frac{d\alpha}{d\phi},        \\
         & = \frac{\phi \sigma^2 \sin(2d\alpha) +
            2d\sinh(\sigma^2\alpha \phi)}{-\alpha \sigma^2 \sin(2d\alpha) +
            2d\cos(2d\alpha)+2d\cosh(\sigma^2\alpha \phi)}.
        \label{trajphi}
    \end{split}
\end{equation}

\begin{figure}[ht]
    \centering
    \includegraphics[width=0.45\textwidth]{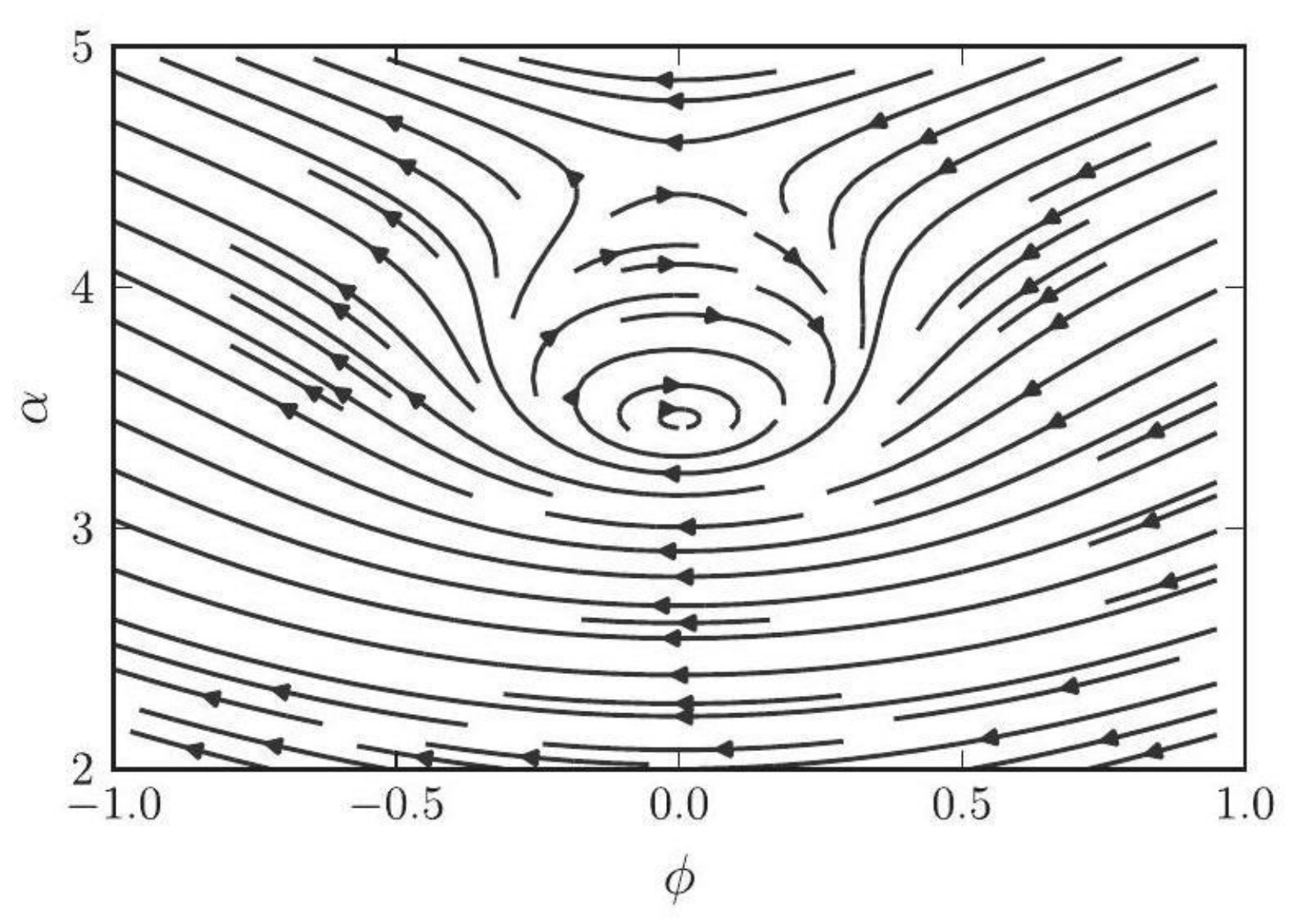}
    \caption{Phase space solutions for the system of Eqs.~\eqref{eq:bohm-planar}
        for $d=-1$ and $\sigma=1$. We notice bouncing solutions and cyclic solutions.
        Figure taken from reference \cite{Bacalhau:2017hja}.}
    \label{fig: Phase space.}
\end{figure}

Figure~\ref{fig: Phase space.} depicts the phase space formed by $\alpha$ and $\phi$, restricted to solutions with a classical limit. All such solutions have a bounce, occurring along the line $\phi=0$. The
classical limit is recovered for large $\alpha$ (i.e., $X \approx\pm 1$ for large
$\alpha$, as seen in Eq.~\eqref{trajphi}). Although cyclic solutions also exist, they are
not physical because they do not contain a classical limit.

For any solution of the Wheeler-De Witt equation, the guidance equations enable us to
find the Bohmian trajectories that describe the evolution of the system. We observe that
Bohmian scale factor solutions that contain a classical limit do not exhibit
singularities, featuring a bouncing point that connects the contracting phase to the
expanding phase.

For the numerical calculations, we use the time $\tau$ defined in
\cite{Bacalhau:2017hja} as
\begin{align}
    \alpha = \alpha_b + \frac{\tau^2}{2},
\end{align}
where the bounce occurs at $\tau=0$. This choice of time is computationally more
well-suited than cosmic or conformal time since those vary over many orders of magnitude
during the evolution of the model. The lapse function $N$ for this choice of time is
\begin{equation}
    N = \frac{\tau}{H}.
\end{equation}
Where $H\equiv \dot{a}/a$ is the Hubble parameter, while $H_0$ will denote its value
today.

We calculate the classical limits of Eqs.~\eqref{eq:bohm-planar} for large values of
$\alpha$, specifically at the points where the hyperbolic function dominates. This
calculation yields the following relations:
\begin{align*}
    \begin{aligned}
        X               & \approx\coth \left(\sigma^{2} \alpha \phi\right),                                                                  \\
        \frac{H}{H_{0}} & \approx \frac{R_{H}}{l_p} \frac{d \mathrm{e}^{-3 \alpha}}{\operatorname{coth}\left(\sigma^{2} \alpha \phi\right)}, \\
        l_p \dot{\phi}  & \approx d \mathrm{e}^{-3 \alpha}.
    \end{aligned}
\end{align*}

In this approach, Eqs.~\eqref{eq:bohm-planar} describe the behavior of any
minisuperspace model containing a canonical scalar field, regardless of its potential,
since the kinetic term generally dominates around the bounce. This dominance allows for
a connection to any classical canonical scalar field with generic potentials, as long as
the potential becomes negligible near the bounce.  

Reference~\cite{Bacalhau:2017hja}
investigates a canonical scalar field model with exponential potential
which connects the classical behaviour presented in Ref.~\cite{Heard:2002} with the quantum evolution described in Ref.~\cite{colistete2000gaussian}. The model exhibits several interesting features: scalar and tensor perturbations consistent
with CMB observations, a kinetic dominated quantum bounce occurring far enough from the Planck scale in order to avoid
compromising the Wheeler-DeWitt approach, and a transient dark energy phase during the
late expansion. As both the quantum and classical phases present the same stiff matter behaviour at high curvatures, where the potential is negligible, they can be joined together, yielding a complete model contracting from an asymptotically dust-dominated phase in the far past, passing through a quantum bounce when the scalar field has a stiff matter behaviour, and expanding to an era where the scalar field behaves as dark energy, when the potential becomes significant. In order to have these properties,
the parameters $d$, $\alpha_b$, $\sigma$, and
$\chi_b$ ($\chi_b$ controlling when the scalar field behaves as dark energy in the
expanding phase, see \cite{Bacalhau:2017hja} for details) must lie within a specific
domain.

\begin{figure}[ht]
    \includegraphics{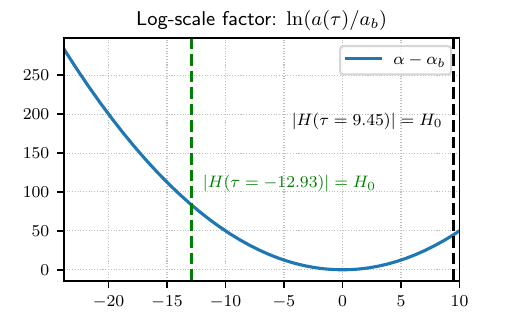}
    \includegraphics{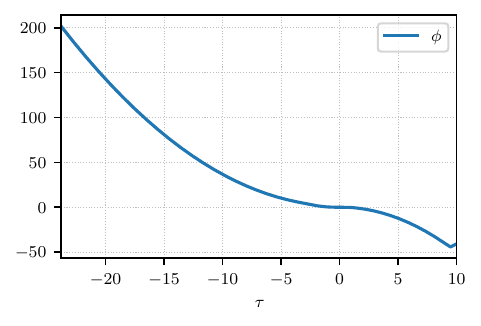}
    \caption{Evolution of the number of e-folds $\alpha - \alpha_b$ with $\tau$. The
        green vertical line marks time during the contraction phase when the Hubble
        radius is equal to the current Hubble radius. The black vertical line indicates
        the same moment during the expanding phase. The scalar field varies within
        the range $\approx (-50, 200)$ over the displayed time interval, encompassing
        most of the relevant scales of the model.}
    \label{fig:alpha}
\end{figure}

In Fig.~\ref{fig:alpha}, we present the evolution of $\alpha$ with respect to $\tau$,
alongside the evolution of the scalar field $\phi$. From this point onward, we evolve
all background and perturbation quantities up to $\tau = 9.3$, which corresponds to a
moment just before the scalar field begins to behave as dark energy, as determined by
the parameters in Table~\ref{modelparam}. This time is selected to align with our
current understanding of the universe, where dark energy currently dominates the energy
budget. Additionally, we limit the evolution to this point to avoid the complexities
associated with the dark energy phase, which would require a more intricate treatment of
perturbations.

In \cite{Bacalhau:2017hja}, we selected two specific sets of parameters, presented in
its Table~\ref{modelparam}, which we also use in this paper. These sets produce
physically motivated models, as discussed in the table legend. Having established the
background model and its key parameters, we now proceed to investigate magnetogenesis
within this framework. All subsequent computations and analyses will be conducted within
this bouncing cosmological setting.

\begin{table}[t]
    \begin{ruledtabular}
        \begin{tabular}{ccccc}
                           & d                   & $\sigma$ & $\alpha_b$            & $\chi_b$          \\
            \midrule Set-1 & $-9 \times 10^{-4}$ & $9$      & $8.3163\times10^{-2}$ & $2\times 10^{36}$ \\
            \midrule Set-2 & $-9 \times 10^{-4}$ & $100$    & $7.4847\times10^{-3}$ & $4\times 10^{36}$ \\
        \end{tabular}
    \end{ruledtabular}
    \caption{Parameter sets for the scalar field-driven bouncing cosmological model that
        yield primordial perturbations consistent with CMB observations. These
        parameters determine the scalar amplitude $\Delta_{\zeta_k}$, the scalar
        spectral index, and the tensor-to-scalar ratio $r$ at horizon crossing. Set-1
        produces $\Delta_{\zeta_k} = 1.4 \times 10^{-10}$ and $r = 1.9 \times 10^{-7}$,
        while Set-2 results in $\Delta_{\zeta_k} = 4.6 \times 10^{-11}$ and $r = 1.3
            \times 10^{-5}$. These values match constraints from CMB data, ensuring the
        viability of the model for early universe predictions. For more details, see
        \cite{Bacalhau:2017hja}.}
    \label{modelparam}
\end{table}

\section{The Electromagnetic Sector}
\label{emsector}

\subsection{Equations of Motion and Power Spectra}

To break the conformal invariance of Maxwell electromagnetism, we introduce a coupling
between the electromagnetic field and a scalar field, described by the Lagrangian
\begin{equation}
    \label{lagr}
    {\cal L}= -f(\phi) F_{\mu\nu}F^{\mu\nu} \;,
\end{equation}
where $f(\phi)$ represents the coupling function. In our analysis, we consider three specific forms: the Gaussian coupling $f_G(\phi)$, the Cauchy coupling $f_C(\phi)$, and the conformal case $f_0(\phi)$, given by:
\begin{align}
    \label{coupling}   f_G(\phi) & = \frac{1}{4} + e^{(\gamma_G^2-\phi^2)/\betaG^2}             \\ &  = \frac{1}{4} + e^{\alphaG^2-(\phi/\betaG)^2} \\
    f_C(\phi)                    & = \frac{1}{4} + \frac{e^{\alphaC^2}}{1 + (\phi/\betaC)^2}, \\
    \label{nocoupling} f_0(\phi) & = \frac{1}{4}.
\end{align}
Here, $\alpha_i$ controls the amplitude and $\beta_i$ sets the scale for $i =
    \mathrm{G}, \mathrm{C}$, both treated as free parameters. For the Gaussian coupling,
the transition between conformal and non-conformal regimes occurs at $\phi =
    \pm \gamma_G=\alphaG\betaG$. The behavior of the Cauchy coupling mirrors the Gaussian one when
$\phi/\betaC \ll 1$, but for $\phi/\betaC \gg 1$, the Cauchy function decays more
slowly, following a $1/\phi^2$ pattern, whereas the Gaussian decays exponentially.
In Fig.~\ref{fig:alpha}, we observe that the scalar field $\phi$ varies between
approximately $-50$ and $200$ over the relevant time interval. Consequently, the
suppression in the Cauchy coupling remains modest, reaching at most
$(\betaC/200)^2$. In order to achieve a significant suppression, a very small $\betaC$ is
required. Therefore, we set $\alphaC = \alphaG$ and $\betaC = \alphaG\betaG
    e^{-\alphaG^2/2}$ so that both the Cauchy and Gaussian couplings cross into the
conformal regime at the same value of $\phi$ (assuming that $e^{\alphaC^2}\gg 1$). This choice of parameters aligns the
two couplings for a consistent comparison, that is, $f_G(\alphaG\betaG) \approx
    f_C(\alphaG\betaG) \approx 1$.

The electromagnetic field equations take the form
\begin{equation}
    \label{eqmov}
    \partial_\mu(\sqrt{-g}\:f\:F^{\mu\nu})=0 \;,
\end{equation}
where $F_{\mu\nu}=\partial_\mu A_\nu-\partial_\nu A_\mu$, and $A_\mu$ represents the
gauge potential. We work in the Coulomb gauge relative to the cosmic time foliation,
setting $A_0=0$ and $\partial_i A^i=0$.

To quantize the electromagnetic field, we expand the operator associated with the
spatial part of the vector potential as follows
\begin{equation}
    \begin{split}
        \label{decomp}
         & \hat{A}_i(t,\mathbf{x})=                        \\
         & \sum_{\sigma=1,2}\int \frac{d^3k}{(2\pi)^{3/2}}
        \left[\epsilon_{i,\sigma}(\mathbf{k})\hat{a}_{\mathbf{k},\sigma}A_{k,\sigma} (t)e^{i\mathbf{k}\cdot\mathbf{x}}+
        \mathrm{H.C.}\right],
    \end{split}
\end{equation}
where $\epsilon_{i,\sigma}(\mathbf{k})$ are the two orthonormal and transverse
polarization vectors associated with the Coulomb gauge, remaining constant across
spatial slices. The comoving wave vector $\mathbf{k}$ has a magnitude given by $k$, and
H.C. denotes the Hermitian conjugate. The operators $\hat{a}_{\mathbf{k},\sigma}$ and
$\hat{a}^\dagger_{\mathbf{k},\sigma}$ are the annihilation and creation operators,
respectively, which satisfy the following commutation relations:
$$
    [\hat{a}_{\mathbf{k},\sigma}, \hat{a}^\dagger_{\mathbf{k'},\sigma'}] =
    \delta_{\sigma\sigma'}\delta(\mathbf{k} - \mathbf{k'}),
$$
and
$$
    [\hat{a}^\dagger_{\mathbf{k},\sigma}, \hat{a}^\dagger_{\mathbf{k'},\sigma'}]=0\;
    ;\;[\hat{a}_{\mathbf{k},\sigma}, \hat{a}_{\mathbf{k'},\sigma'}]=0.
$$
The time-dependent coefficients $A_{k,\sigma}(t)$ and their corresponding momenta
$\Pi_{k,\sigma} \equiv 	4af{\dot{A}}_{k,\sigma}(t)$ must satisfy the vacuum
normalization
\begin{equation}\label{vacuumnorma}
    A_{k,\sigma}(t)\Pi^*_{k,\sigma}(t)
    -A^*_{k,\sigma}(t)\Pi_{k,\sigma}(t) = i,
\end{equation}
for each $k$ and $\sigma$.

As it is well known, the gauge-fixed electromagnetic field, in the absence of charges, is
equivalent to that of two free real scalar fields, corresponding to each polarization
direction $\sigma$. Since we are considering an isotropic background, we choose a single
time-dependent coefficient to describe both polarizations, i.e. $A_{k,1} = A_{k,2}
    \equiv{A_{k}} $, which yields the same vacuum for both cases.

Inserting this decomposition into the equations of motion Eq.~\eqref{eqmov} leads to the
following equations for the modes $A_k(t)$:
\begin{align}
    \label{pot}
    \ddot{A}_k+\left(\frac{\dot a}{a}+\frac{\dot f}{f}
    \right)\dot{A}_k +\frac{k^2}{a^2}A_k=0 \;.
\end{align}
where in the equation above, dots denote derivatives with respect to cosmic time. We now
introduce a set of dimensionless quantities that simplify the analysis
\begin{align}
    Y   & \equiv \frac{a}{a_0}, \quad  A_{sk} \equiv \frac{A_k}{\sqrt{R_{H}}}, \\
    x_b & \equiv \frac{a_0}{a_b} \;, \quad k_s \equiv \frac{k R_{H_0}}{a_0},
    \quad \eta_s \equiv \int \frac{\textup{d}t}{R_{H_0}Y},
\end{align}
where $R_H \equiv R_{H_0}/a_0$ is the co-moving Hubble radius, and $R_{H_0}$ is the
Hubble radius today. Therefore, Eq.~\eqref{pot} turns into
\begin{align}
    \label{potetas}
    A_{sk}^{\prime \prime}+\frac{f^{\prime}}{f}A_{sk}^{\prime}+k_s^2 A_{sk} = 0,
\end{align}
where now the prime indicates a derivative with respect to $\eta_s$. 

\begin{figure}[ht]
    \includegraphics{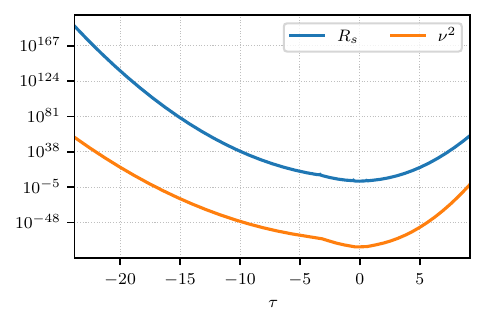}
    \caption{Evolution of the frequency $\nu$ and the Ricci scale with $\tau$.}
    \label{fig:rs_nu}
\end{figure}

Generally, the Hamiltonian
generating the general $\tau$ evolution ($\dd t=N \dd\tau$) of such systems reads
\begin{equation}
    \label{hamiltonian}
    \mathcal{H} = \frac{\Pi_{sk}^2}{2m} + \frac{m\nu^2 A_{sk}^2}{2} \;,
\end{equation}
It is equivalent to a Hamiltonian describing a harmonic oscillator with time dependent effective mass $m = af/N$
and frequency $\nu = Nk_s/a$. We plot the frequency $\nu$ along the Ricci scale $l_R\equiv R^{-1/2}$, where $R$ is the Ricci scalar, in
Fig.~\ref{fig:rs_nu}. Note that, given our choices of dimensionless variables, all
variables with units of distance are expressed in units of the Hubble radius today.

The energy densities of the electric and magnetic fields are respectively given by:
\begin{align}
    \rho_E & =\frac{f}{8\pi}g^{ij}\dot{A}_i \dot{A}_j\;,                                                     \\
    \rho_B & =\frac{f}{16\pi}g^{ij}g^{lm}(\partial_j A_m-\partial_m A_j) (\partial_i A_l-\partial_l A_i) \;,
\end{align}
where $g^{ij}=\delta^{ij}/a^2$ represents the spatial components of the inverse metric.
To derive the spectral energy densities, we insert expansion \eqref{decomp} into
$\rho_E$ and $\rho_B$, thus transforming $\hat{\rho}_E$ and $\hat{\rho}_B$ into quantum
operators and computing their vacuum expectation values:
\begin{align}
    \bra{0} \hat{\rho}_B \ket{0} & = \frac{f}{2\pi^2 R_{H_0}^4 Y^{4}} \int \dd{\ln{k_s}} \;  \vert A_{sk}\vert^2 k_s^5, \label{eq:mag-density}          \\
    \bra{0} \hat{\rho}_E \ket{0} & = \frac{f}{2\pi^2 R_{H_0}^4 Y^{4}} \int \dd{\ln{k_s}} \; \vert A^{\prime}_{sk}\vert^2 k_s^3. \label{eq:elec-density}
\end{align}
Defining the spectra as
\begin{align}
    {{\cal P}_{i}} \equiv \frac{\textup{d}\bra{0} \hat{\rho}_i \ket{0}}{\dd{\ln{k}}} \;,  \quad i=E,B \;,
\end{align}
we find:

\begin{align}
    {\cal P_{B}} & \equiv B^2_\lambda = \frac{f}{2\pi^2 R_{H_0}^4}\frac{\vert A_{sk}\vert^2}{Y^4}k_s^5,
    \label{magpow}                                                                                                                                                                  \\
    {\cal P_{E}} & \equiv E_{\lambda}^2=\frac{f}{2\pi^2 R_{H_0}^4}\frac{\vert A_{sk}^\prime\vert^2}{Y^4}k_s^3 = \frac{1}{2\pi^2 R_{H_0}^4}\frac{\vert \Pi_{sk}\vert^2}{f Y^4}k_s^3.
    \label{elecpow}
\end{align}
In the last line, we also expressed ${\cal P_{E}}$ in terms of the momentum canonically
conjugate to the gauge field, $\Pi_{sk} = f A_{sk}^\prime$, which is equivalent to the
electric field mode itself. Apart from a potential time-dependent coupling, note that
both power spectra exhibit the same $Y^{-4}$ behavior, as expected for an
electromagnetic field. Any departure from this behavior must arise from the function $f$
and the magnitude of the field modes, which only vary if the coupling is not constant.

Finally, we express the magnetic and electric fields, $B_{\lambda}$ and $E_{\lambda}$,
using $H_0^2 \approx 1.15 \times 10^{-64}$ G. This conversion from the Hubble radius to Gauss
is done to facilitate comparisons with the existing literature, which commonly uses
Gauss as the unit of measurement.
\begin{align}
    \label{est}
    B_\lambda & =\sqrt{\frac{f}{2\pi^2}}
    \frac{\vert A_{sk}\vert }{Y^2}k^{5/2}\:1.15\times 10^{-64} {\rm G}, \\
    E_\lambda & =\sqrt{\frac{1}{2 \pi^2 f}}
    \frac{\vert  \Pi_{sk}\vert }{Y^2}k^{3/2}\:1.15\times 10^{-64} {\rm G}.
\end{align}

\subsection{Vacuum Initial Conditions}

In the context of a cosmological bouncing model driven by a scalar field, we define
vacuum initial conditions to accurately determine the behavior of the electromagnetic
field throughout the bounce and the subsequent expansion phase. Our model considers two
types of non-minimal couplings between the electromagnetic field and the scalar field:
Gaussian and Cauchy couplings. These couplings may influence the initial conditions and the
subsequent evolution of the field modes.

We establish the adiabatic vacuum initial conditions to ensure that each mode of the
electromagnetic field starts in its lowest possible energy state within a smoothly
varying background. This is a very reasonable assumption, as in the asymptotic past of such models the universe is almost empty and flat, with almost nothing but quantum fluctuations of an almost Minkowski vacuum state for fields and cosmological perturbations. By initializing the field in a vacuum state, we ensure that any
electromagnetic fields observed in the expansion phase can be attributed solely to the
dynamics of the contraction and bounce, without interference from initial modes. These
initial conditions are mathematically expressed in terms of the coupling function
$f(\phi(\eta_s))$, where $\phi$ is the time-dependent scalar field in conformal time
$\eta_s$.

For the electromagnetic field $A_\mu$ in a cosmological background, the leading order
adiabatic vacuum initial conditions are specified as follows:
\begin{equation}\label{adiabatic1}
    A_k(\eta_s) = \frac{e^{-ik\eta_s}}{\sqrt{2k f(\phi(\eta_s))}},
\end{equation}
\begin{equation}\label{adiabatic2}
    \Pi_k(\eta_s) = f(\phi(\eta_s)) A'_k(\eta_s) = -i\sqrt{\frac{f(\eta_s) k}{2}} e^{-ik\eta_s}.
\end{equation}
These expressions guarantee that, in the distant past (before the bounce), when the
coupling function $f(\phi) \approx \frac{1}{4}$, the modes take the form
\begin{equation}
    |A_k| = \sqrt{\frac{2}{k}},
\end{equation}
\begin{equation}
    |\Pi_k| = \sqrt{\frac{k}{8}}.
\end{equation}

Generically, the adiabatic vacuum initial conditions can be derived using the complex
structure approach presented in Ref.~\cite{Penna-Lima:2022dmx}. In this framework, based on the physical idea that the effective frequency of the mode varies much slower than the time scale defined by the frequency itself, the adiabatic vacuum
initial conditions are expressed as functions of
\begin{align}
    F_n = \left(\frac{1}{2\nu} \frac{\textup{d}}{\textup{d}t}\right)^n \xi \;,
\end{align}
where $1 \gg F_1 \gg F_2 \gg F_n$, and $\xi = \ln(m\nu)$. For our model, we set $t =
    \eta_s$, $\nu = k_s$, and $m = f$. This condition implies that the function $\xi = \ln(f
    k_s)$ changes slowly compared to $\int 2 k_s \, \text{d}\eta_s$. In our numerical
analysis, we utilize the adiabatic vacuum initial conditions up to the fourth order,
applying this approximation until the truncation error becomes significant. 

To illustrate how the couplings satisfy the adiabatic vacuum initial conditions as
$\eta_s \rightarrow -\infty $, we examine their asymptotic behavior. In this limit, the
scalar field $ \phi(\eta_s) $ typically grows large, as seem in Fig.~\ref{fig:alpha}.
For the Gaussian coupling, the exponential term $ e^{-\phi^2/\beta^2} $ quickly tends to
zero, leading to $ f_G(\phi(\eta_s)) \approx \frac{1}{4} $. Similarly, for the Cauchy
coupling, the term $ \frac{1}{1 + (\phi/\beta)^2} $ also approaches zero, resulting in
$f_C(\phi(\eta_s)) \approx \frac{1}{4}$. Thus, in the limit \(\eta_s \rightarrow -\infty\), both coupling functions converge to a
constant value of \(\frac{1}{4}\), simplifying the initial conditions to:
\begin{equation}
    A_k(\eta_s) \approx \sqrt{\frac{2}{k}} e^{-ik\eta_s},
\end{equation}
\begin{equation}
    \Pi_k(\eta_s) \approx -i\sqrt{\frac{k}{8}} e^{-ik\eta_s}.
\end{equation}

By studying these two extremes -- the sharply decreasing Gaussian and the more gradually
decreasing Cauchy coupling -- we can effectively map the general behavior of such
couplings. These represent limiting cases within a broader family of decreasing
functions, allowing us to capture a wide range of possible dynamics and better
understand the influence of different coupling behaviors on the evolution of the
electromagnetic field in the model.

To accurately capture the dynamics of the electromagnetic field, we implement the
adiabatic vacuum initial conditions specific to each coupling function within our
numerical simulations. By initializing the field modes in their appropriate vacuum
states and evolving them through the bounce, we ensure that the resulting predictions,
including the magnetic field power spectrum, are robust. This method eliminates
potential artifacts that could arise from improper initial conditions, while maintaining
precision up to the point where the adiabatic approximation breaks down due to
truncation errors.

\section{Numerical Analysis of Perturbations}
\label{numeric}
\subsection{Modes}

We examine the time evolution of the magnetic and electric modes, $|A_{sk}|^2$ and
$|\Pi_{sk}|^2$, focusing on how the scale parameter $\betaG$ affects the magnetic modes
for $\alphaG = 13.15$ and $\alphaG = 14.4$. The values of $\betaG$ in
Table~\ref{couplingparam} determine the transition point between the conformal and
non-conformal regimes, defined by $\phi_0 = \alphaG \betaG$, which ranges from
approximately 10 to 50. As shown in Fig.~\ref{fig:alpha}, this corresponds to the time
interval $\tau \in [-10, -5]$.

\begin{figure*}[ht]
    \begin{center}
        \includegraphics[scale=0.9]{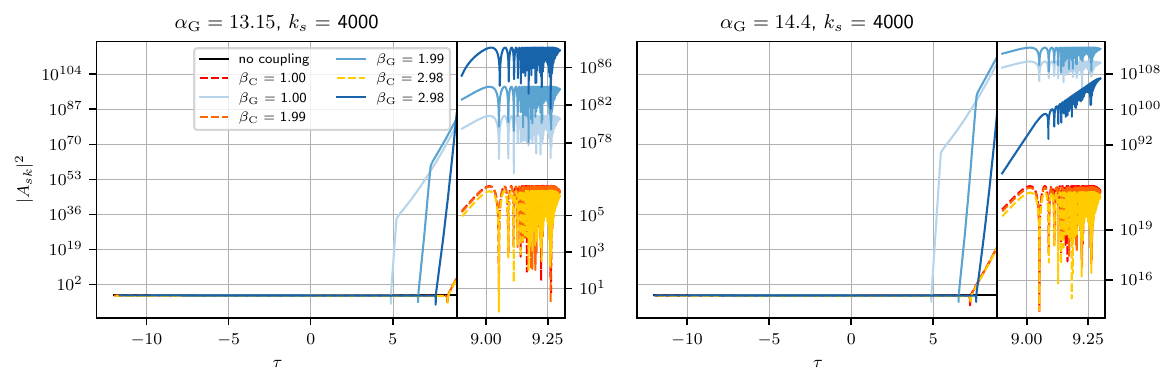}
    \end{center}
    \caption{Time evolution of magnetic modes $\vert A_{sk} \vert^2$. Warm gradient curves (red, orange, yellow)          represent modes with a Cauchy coupling, while blue curves represent modes with a
        Gaussian coupling. The black curve corresponds to the uncoupled case. The
        magnetic modes only begin to evolve significantly during the expansion phase
        ($\tau > 0$). For both couplings, the amplitude of the modes increases with
        $\alphaG$, with the amplification occurring as the modes become super-Hubble in
        the expanding phase. We zoom in on the region $8.9<\tau<9.3$ to display the onset of oscillations corresponding to the modes entering the Hubble radius.}
    \label{J11}
\end{figure*}

\begin{figure*}[ht]
    \begin{center}
        \includegraphics[scale=0.9]{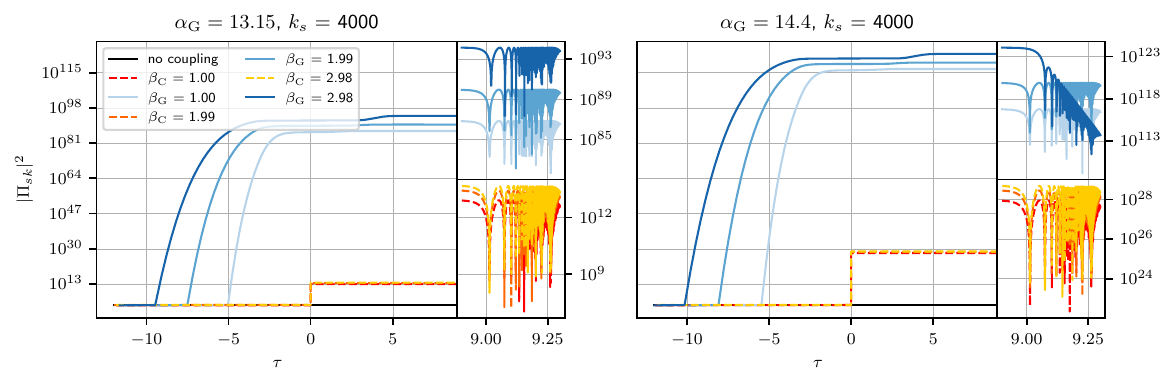}
    \end{center}
    \caption{Time evolution of electric modes $\vert \Pi_{sk} \vert^2$. Warm gradient curves (red, orange, yellow)          represent modes with a Cauchy coupling, and blue curves represent modes with a
        Gaussian coupling. The black curve corresponds to the uncoupled case. The
        electric modes start evolving in the contracting phase ($\tau < 0$). For the
        Cauchy coupling, the momentum amplitude is enhanced relative to the no-coupling
        case near the bounce. The Gaussian modes follow a similar pattern but with
        varying amplification timescales. Both couplings yield larger amplitudes as
        $\alphaG$ increases. We zoom in on the region $8.9<\tau<9.3$ to display the onset of oscillations corresponding to the modes entering the Hubble radius.}
    \label{J22}
\end{figure*}

We begin by plotting these modes because, in the absence of a non-trivial coupling,
their magnitudes should remain constant, providing a baseline for comparison.
Figure~\ref{J11} illustrates the impact of $\betaG$ on the magnetic mode, while
Fig.~\ref{J22} demonstrates similar effects on the electric modes. Both figures focus on
a fixed wave-number, $k = 4000$ (approximately 1 Mpc), although the overall behavior is
consistent across larger scales. The uncoupled case, represented by the black line,
serves as a reference throughout the analysis. Table~\ref{couplingparam} provides a
summary of the parameter ranges used in this study.

\begin{table}[t]
    \begin{ruledtabular}
        \begin{tabular}{ccc}
            Parameter & $\alphaG$                   & $\betaG$              \\
            \midrule
            Range     & $\left[13.15, 15.18\right]$ & $\left[1,2.99\right]$ \\
        \end{tabular}
    \end{ruledtabular}
    \caption{Range of coupling parameters $\alphaG$ and $\betaG$.}
    \label{couplingparam}
\end{table}

We first observe that while the Cauchy coupling evolves over time, its evolution occurs
on a much smaller time scale compared to the Gaussian coupling, rendering its effects
nearly imperceptible, partially because one needs a very small $\betaC$ in order to reach the conformal limit before the standard cosmological model evolution in the early universe begins to take place.
Although the coupling parameters do influence the amplitude of the modes, the Cauchy
coupling (depicted by the warm gradient curves) fails to generate notable magnetic or electric
fields. While some amplification occurs, it remains insufficient to produce observable
effects. For the same $\alphaG$, the Cauchy coupling results in effects that are many
orders of magnitude smaller than those of the Gaussian coupling. One could increase
$\alphaC$ further to amplify the effects, but doing so would require an even smaller
$\betaC$ to maintain the coupling within the conformal regime. We will explore these
effects in more detail in Figs.~\ref{PB} and \ref{PE}.

Second, we observe that modes with Gaussian coupling (shown by the blue curves) evolve
similarly to uncoupled modes in the distant past, as expected, given that the coupling
is approximately constant in this regime. During the contraction phase, only the
electric modes $\vert \Pi_{sk} \vert^2$ increase, while the magnetic modes $\vert A_{sk}
    \vert^2$ remain nearly constant. It is only during the expansion phase that the magnetic
modes begin to grow. This behavior mirrors what is observed in other bouncing models
coupled to gravity, as discussed in \cite{Frion:2020bxc}. A broader Gaussian coupling
(with larger $\beta$, indicated by the darker shades of blue) affects modes farther from
the bounce, further amplifying the contrast between electric and magnetic mode
evolution. This outcome is expected since larger $\beta$ values cause the coupling to
transition to the non-conformal regime earlier, allowing the modes to amplify sooner.

The behavior of the modes can be understood through the evolution of the Hamiltonian
equations derived from Eq.~\eqref{hamiltonian}:
\begin{subequations}
    \begin{align}
        \dot{A}_{sk}   & = \frac{\Pi_{sk}}{m}, \\
        \dot{\Pi}_{sk} & = -m\nu^2 A_{sk}.
    \end{align}
\end{subequations}
In the super-Hubble regime, as the coupling increases, the effective mass $m$ rises,
leading to an increase in $\Pi_{sk}$ while $A_{sk}$ remains approximately constant. When
the mode transitions into the sub-Hubble regime during the expanding phase, the
increased momentum is transferred to the mode amplitude. This dynamic explains the
behavior observed in Figs.~\ref{J11} and \ref{J22}. The key difference between the
couplings lies in the timing of this variation: for the Cauchy coupling, it occurs very
close to the bounce, whereas for the Gaussian coupling, the variation is spread over a
longer period, due to the large difference between $\betaC$ and $\betaG$, allowing for
more gradual amplification.

\subsection{Power Spectra}

\begin{figure*}[ht]
    \begin{center}
        \includegraphics[scale=0.9]{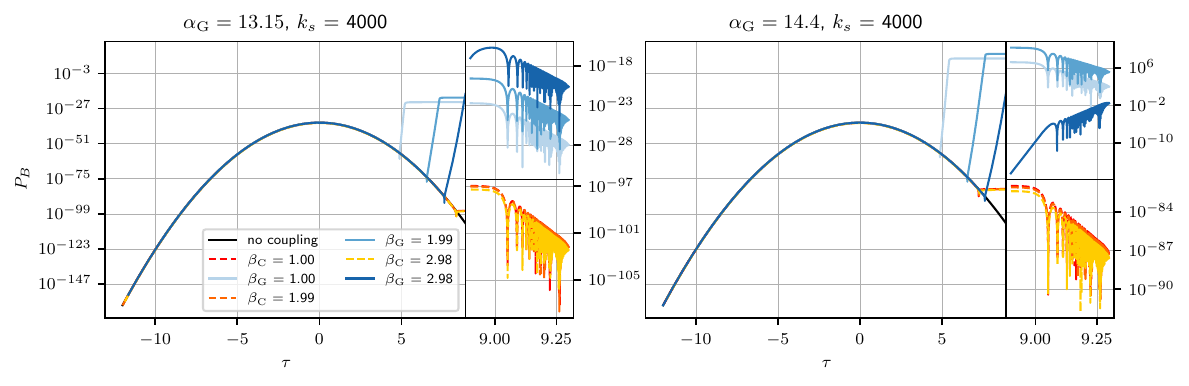}
    \end{center}
    \caption{Evolution of the magnetic power spectrum $P_B$ with time for modes
        $k=4,000$. The colors and parameters are the same as for the modes in figures
        \ref{J11} and \ref{J22}. A Cauchy coupling is unable to explain the origin or
        large-scale magnetic fields, while a Gaussian coupling is an efficient way of
        producing strong seed magnetic fields.}
    \label{PB}
\end{figure*}

Based on the previous analysis, we now investigate the dependence of the magnetic and
electric power spectra, $P_B$ and $P_E$, on the parameters $\alphaG$ and $\betaG$. As shown in
Fig.~\ref{PB}, the magnetic power spectra for the Cauchy modes exhibit minimal
amplification, beginning near the baseline and increasing only marginally during the
expansion phase. For $\alphaG = 13.15$ (left panel), the difference is a few orders of
magnitude, rising to about 10 orders of magnitude for $\alphaG = 14.4$ (right panel).
Consequently, the Cauchy coupling proves ineffective in generating primordial magnetic
fields with significant amplitudes. In contrast, the Gaussian modes undergo substantial
growth during the expansion phase, amplifying their magnetic power spectra to levels
capable of explaining the observed magnetic fields on very large scales. In the
following sections, we will further constrain the model's parameter space to explore the
origins of these fields.

\begin{figure*}[ht]
    \begin{center}
        \includegraphics[scale=0.89]{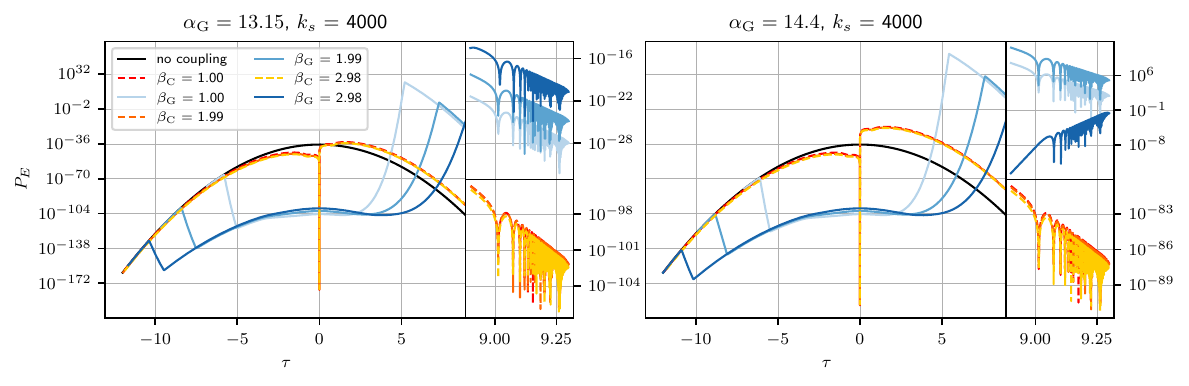}
    \end{center}
    \caption{Evolution of the electric power spectrum $P_E$ with time for modes
        $k=4,000$. The colors and parameters are the same as for the modes in figures
        \ref{J11} and \ref{J22}. Both the Cauchy and Gaussian couplings see their amplitude decreasing
        during contraction ($\tau<0$) and increasing during expansion ($\tau>0$), though the amplification is much stronger in the Gaussian case.}
    \label{PE}
\end{figure*}

While the electric power spectra for both Cauchy and Gaussian modes exhibit similar
qualitative behaviors, they differ in scale and amplitude. Specifically, both power
spectra decrease during contraction ($\tau<0$) and increase during expansion ($\tau>0$).
This is illustrated in Figure~\ref{PE}, where the primary distinction between the two
types of couplings is their overall magnitude rather than the general trend of the
mode evolution.

\subsection{Backreaction}

Similar to magnetic power spectra, the electric power spectra dominate only
significantly after the quantum-dominated regime around the bounce. Consequently, both
electric and magnetic energy densities become important only in the classical regime,
which leads us to conclude that no backreaction occurs during the quantum regime.

\subsection{Parameter Space}

\begin{figure*}[ht]
    \includegraphics{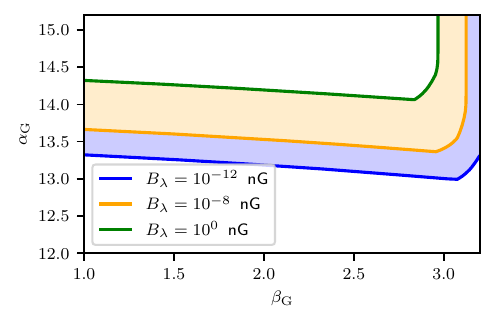}
    \includegraphics{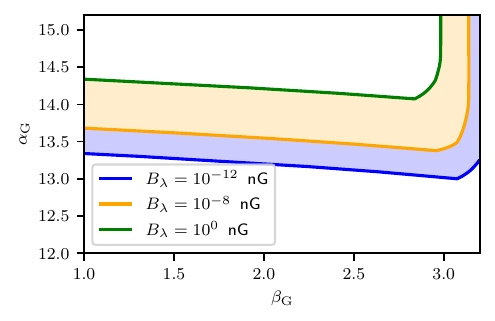}
    \caption{Viable parameter space $(\alphaG, \betaG)$ for the generation of primordial
        magnetic fields compatible with observations. The parameter space for Set-1 is
        on the left, for Set-2 on the right. The green curve corresponds to the upper
        constraints on magnetic field amplitudes $\approx 1$ nG, the orange curve to
        lower constraints from $\gamma$-rays, and the blue curve a lower threshold to
        initiate dynamo amplification. The orange region contains values of $\alphaG$
        and $\betaG$ leading to viable magnetic seed fields, and overlaps with the
        dynamo (blue) region. For both sets, there is a sharp increase in the value of
        $\alphaG$ needed to obtain a magnetic field amplitude of 1 nG when $\betaG \to
            3$.}
    \label{paramspace}
\end{figure*}

In Fig.~\ref{paramspace}, we illustrate the parameter space where our model produces
viable seed magnetic fields. The green curve represents the upper bound of approximately
1 nG on the amplitude of seed magnetic fields, as constrained by various surveys,
including observations from the CMB~\cite{Ade:2015cva}, Ultra-High-Energy Cosmic Rays
(UHECR) \cite{Bray:2018ipq}, and 21-cm hydrogen lines \cite{Minoda:2018gxj}. The orange
curve denotes the lower bound derived from the non-detection of secondary GeV
$\gamma$-rays around TeV blazars \cite{Taylor:2011bn,MAGIC:2022piy}, though this limit
remains debated within the astrophysical community \cite{Broderick:2011av,
    Subramanian:2019jyd} and should be interpreted with caution. Additionally, the blue
curve represents the theoretical threshold required to initiate a galactic dynamo
\cite{subramanian1994, Martin:2007ue}. For the remainder of this study, we focus
exclusively on the Gaussian coupling due to its viability in generating magnetic fields
within these constraints.

\begin{figure*}[ht]
    \centering
    \includegraphics{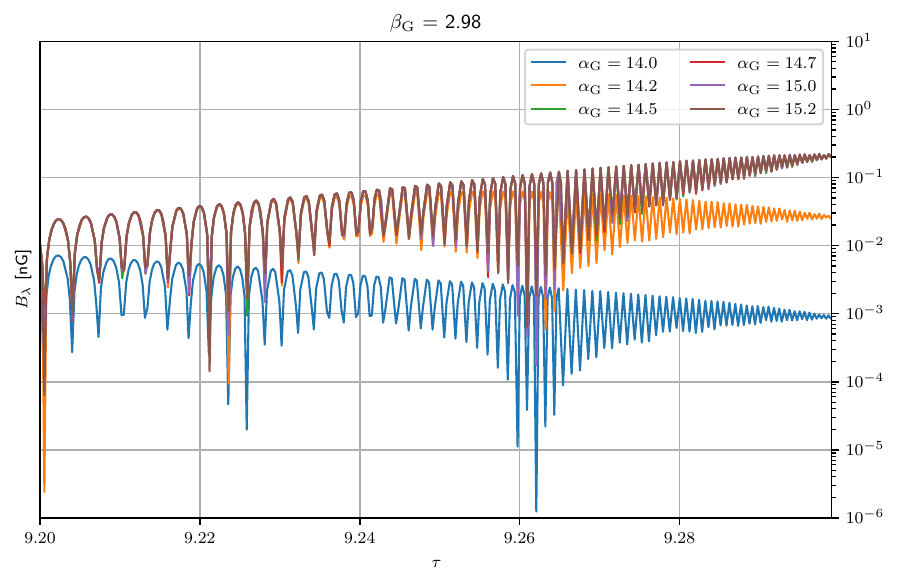}
    \caption{At fixed $\betaG$, increasing $\alphaG$ ends up changing the slope of the
        magnetic field. Starting with a negative slope for lower values of $\alphaG$
        (blue curve) and increasing $\alphaG$ changes the sign of the slope. In this
        example, for $\alphaG > 14.48$, all curves are superposed. Therefore, there is a
        saturation mechanism related to the sign of the slope.}
    \label{powerbreak}
\end{figure*}

The left and right panels of Fig.~\ref{paramspace} display the parameter spaces for
Set-1 and Set-2, respectively. Both cases exhibit a similar pattern, notably a sharp
rise in $\alphaG$ as $\betaG$ nears 3, required to generate a magnetic seed field
amplitude of 1 nG. This behavior is tied to the time evolution of the magnetic field. In
Fig.~\ref{powerbreak}, we illustrate the magnetic field evolution over a brief interval
($9.2 < \tau < 9.3$). By holding $\betaG$ fixed and increasing $\alphaG$ to achieve a 1
nG amplitude today, we see a sign change in the spectrum's slope. After this point,
further increases in $\alphaG$ lead to only marginal changes in $B_{\lambda}$.

\subsection{Amplitude of Magnetic Fields}

\begin{figure*}[ht]
    \begin{center}
        \includegraphics{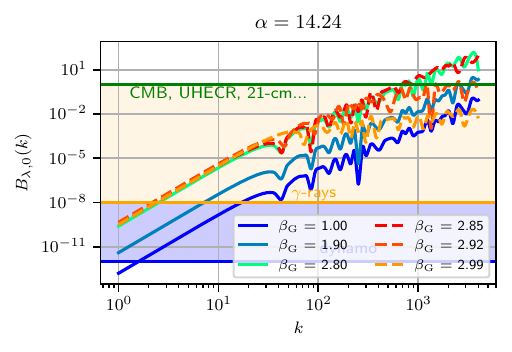}
        \includegraphics{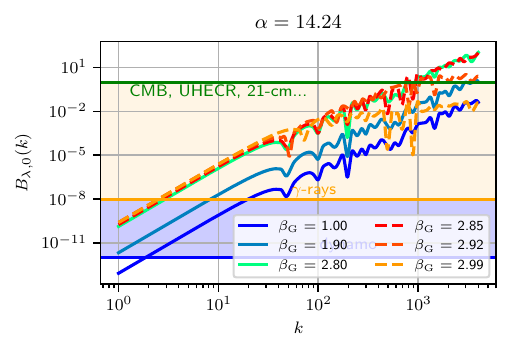}\\
        \includegraphics{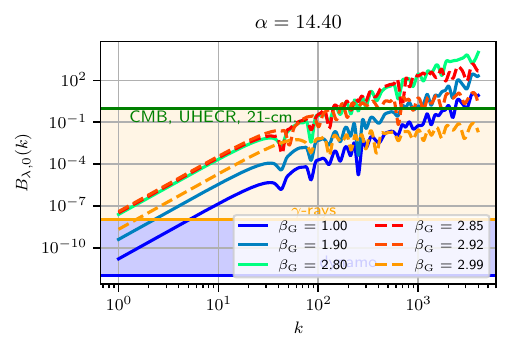}
        \includegraphics{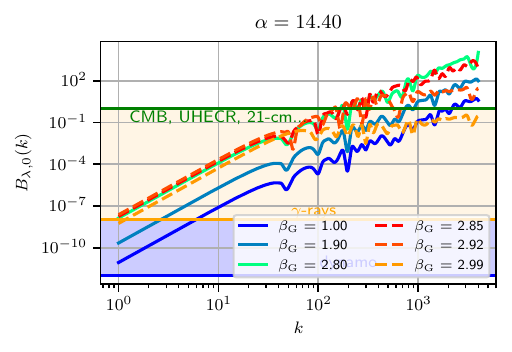}
    \end{center}
    \caption{Evolution of the amplitude of magnetic fields today with the scale. We keep
        the same color code for bounds on the amplitude as in figure \ref{paramspace}.
        Figures for Set-1 are on the left, for Set-2 on the right. For all values of
        $\alphaG$ and $\betaG$, magnetic fields have a blue-tilted spectral index $n_B$.
        On the largest scales $0<k\leq 30$, $n_B=4$ always. For scales $k>30$, we have
        $n_B \simeq 3$ while $1<\beta<2.8$, and $0.5<n_B<3$ for $\beta>2.8$.}
    \label{amplitude}
\end{figure*}

Figure~\ref{amplitude} depicts the present-day amplitude of magnetic fields,
$B_{\lambda,0}$, across different scales. The trends observed are consistent between
Set-1 and Set-2. For super-Hubble modes, specifically in the range $1 < k \leq 30$, the
magnetic fields follow a power-law distribution with a spectral index of $n_B \simeq 4$,
largely independent of the specific values of $\alphaG$ and $\betaG$. However, for modes
within the Hubble radius, the spectral index $n_B$ becomes more sensitive to $\betaG$.
When $\betaG$ is between 1 and 2.8, the average scalar index $n_B$ decreases from 4 to
approximately 3. Beyond this range, particularly for $\betaG > 2.8$, the slope undergoes
a more pronounced shift, dropping to as low as $n_B=0.5$ for $\betaG = 2.99$, though it
remains positive.

\section{Discussion}
\label{discussion}

The results of this study suggest that a bouncing universe
together with a coupling between the scalar field $\phi$ which sources the background model and the
electromagnetic field can lead to sufficient primordial magnetogenesis in accordance with observations. We explored the evolution of the electromagnetic field starting
from the contraction phase, through the bounce, and into the expansion phase, beginning
with an initial adiabatic vacuum state. Our numerical analysis reveals that magnetic
fields on a 1 Mpc scale can achieve amplitudes consistent with current cosmological
observations. This is significant because it shows that, in the presence of the coupling,
the magnetic fields generated during the bounce are sufficiently strong to serve as
seeds for further amplification by dynamo mechanisms in astrophysical structures,
supporting the potential of this model to explain the observed magnetic fields in the
universe.

Two critical parameters in the model are $\alphaG$ and $\betaG$, which play a pivotal
role in shaping the results. The parameter $\betaG$ determines the width of the coupling
functions, while $\alphaG$ sets the coupling amplitude. The value of $\betaG$ controls
the duration of the coupling's influence, with larger values extending the coupling's
effect over a longer period around the bounce. Our analysis shows that a range of
$\betaG$ values between 1 and 2.99 can generate magnetic fields consistent with current
observations. In particular, values near the upper bound, $\betaG \approx 2.99$, yield
stronger magnetic fields. This occurs because higher $\betaG$ values prolong the
interaction between the scalar field $\phi$ and the electromagnetic field, leading to
greater energy transfer to the magnetic field. This not only amplifies the magnetic
field's strength but also smooths and extends the oscillations in the power spectrum
over time.

The parameter $\alphaG$ influences the strength of the coupling between the scalar field
$\phi$ and the electromagnetic field. Larger values of $\alphaG$ result in greater
amplification of the magnetic modes by intensifying the interaction between $\phi$ and
the electromagnetic field. Our results show that for $\alphaG$ in the range of $12.7$ to
$15.18$, the generated magnetic fields align with observational limits. As $\betaG$
approaches 3, higher values of $\alphaG$ are necessary to achieve a magnetic field
amplitude of 1 nG today. This is because a stronger coupling (higher $\alphaG$) provides
additional energy for amplifying the magnetic fields to match cosmological observations.
Moreover, varying $\alphaG$ affects the power spectrum's slope and the degree of
amplification, leading to significant changes in the power spectrum's qualitative
features, such as the prominence of peaks and troughs.

Another important aspect are the oscillations observed at the end of the evolution of the
power spectra. These oscillations begin when the modes enter the Hubble radius and start
behaving like typical electromagnetic waves, as they are no longer super-Hubble. At this
point, the oscillations arise purely from the dynamics of the electromagnetic fields,
rather than interactions with other matter components.

In the context of primordial magnetogenesis in a bouncing universe model, the issue of
backreaction -- i.e., the influence of the generated magnetic fields on the dynamics of
the universe -- does not present itself as a significant problem. The energy density of
the magnetic fields $\rho_B$ must be compared with the background energy density
$\rho_\phi$ to assess the impact of backreaction. In our model, we verify that $\rho_B$
remains significantly smaller than $\rho_\phi$ throughout the entire cosmological
evolution considered. Specifically, the energy density of the generated magnetic fields
$\rho_B$ is given by
\begin{equation*}
    \rho_B = \int_{k_{\text{min}}}^{k_{\text{max}}} \frac{d\ln k}{8\pi} P_B(k),
\end{equation*}
where $P_B(k)$ is the magnetic field power spectrum, calculated numerically. Comparing this
expression with the background energy density $\rho_\phi$, which is given by:
\begin{equation*}
    \rho_\phi = \frac{1}{2} \dot{\phi}^2 + V(\phi).
\end{equation*}
The terms that dominate $\rho_\phi$ depend on the phase of the universe. During the
contraction phase, the kinetic term $\frac{1}{2} \dot{\phi}^2$ tends to dominate, making
$\rho_\phi \propto e^{-6\alpha}$. In the expansion phase, the energy density of the
scalar field is influenced by the contribution of the potential $V(\phi)$, which becomes
more relevant. This dependence of the background energy density on the scale factor,
along with the Planck length, corroborates the stability of the model concerning the
back-reaction problem. The same is true concerning the energy density of electric field.

For example, considering typical values of the potential $V(\phi)$ and the evolution of
the scale factor $a(t)$, the energy density $\rho_\phi$ can vary from $10^{-12}
    \text{GeV}^4$ during the contraction phase to $10^{-9} \text{GeV}^4$ in the expansion
phase. These values are widely accepted in the cosmological literature and show that,
even with the amplification of the magnetic fields, the energy density $\rho_B$ remains
well below $\rho_\phi$, ensuring that backreaction is not a problem in this scenario.

\section{Conclusions}

In this work, we presented the generation of primordial magnetic fields in the context
of a cosmological bounce through a coupling between the electromagnetic field and a
scalar field. We considered a homogeneous and isotropic background filled with a scalar field which behaves as a
pressureless (dark) matter fluid in the asymptotic contracting phase, as a stiff-matter fluid near the bounce, and as dark energy in some period of the
expanding phase. The bounce is produced by quantum effects described using the de
Broglie-Bohm quantum theory, motivated by the inconsistency of
using standard quantum mechanics in quantum cosmology \cite{Pinto-Neto:2013toa}. The background model has scalar cosmological perturbations in accordance with CMB observations, see Ref.~\cite{Bacalhau:2017hja} for details.

Some advantages of the bounce magnetogenesis are the absence of the strong
coupling problem and backreaction. The model is characterized by three main parameters: the coupling
amplitude $\alphaG$, the width of the coupling function $\betaG$, and the specific form
of the coupling (Gaussian or Cauchy), which have been explored, leading to relevant primordial magnetic fields which can source the observed magnetic fields observed in galaxies and clusters. Only the Gaussian coupling, with its intrinsic abrupt fall-off, can yield such magnetic fields. The full electromagnetic field evolves from fluctuations of an adiabatic vacuum properly defined in the far past of the cosmological model either for the
Gaussian and Cauchy couplings. 

The inclusion of theoretical limits on gravitational wave production, see Ref.~\cite{Caprini:2001nb} will be investigated in the future. This will be even more
relevant with the upcoming observations from LISA \cite{Caprini:2009yp, Caprini:2018mtu,
    Saga:2018ont, Pol:2019yex}.

A second point of interest would be to consider other possible backreaction effects. It
has been shown recently that vacuum polarization in a dielectric medium, the so-called
Schwinger effect, increases the conductivity of the medium and subsequently halts the
production of the magnetic field \cite{Sobol:2018djj, Sobol:2019xls, Sharma:2017eps,
    Sharma:2018kgs, Shakeri:2019mnt}. This effect could lead to weaker magnetic fields than
expected and could further constrain our model. It is important to investigate how these
interactions affect the evolution of the magnetic field and whether there are ways to
mitigate these effects to preserve the viability of the bounce magnetogenesis model.

As a possible extension of our work, other non-minimal couplings between the
electromagnetic field and the scalar field can be explored. For example, couplings
dependent on other functional forms of the scalar field or couplings that vary in time
may provide new perspectives on the generation of primordial magnetic fields.

We leave these questions for future work, encouraging the exploration of new models
and the consideration of additional effects that may impact the generation and evolution
of primordial magnetic fields in the context of a cosmological bounce.

\section*{Acknowledgments}
NPN acknowledges the support of CNPq of Brazil under grant PQ-IB
310121/2021-3. SDPV acknowledges the support of CNPq of Brazil under grant PQ-II 316734/2021-7.

\bibliographystyle{plainnat}
\bibliography{main.bib}

\end{document}